\documentclass[
prl,
twocolumn,
nofootinbib,
floatfix,
showpacs ,amssymb, aps]{revtex4}
\usepackage{amsmath,slashed}

\usepackage{color}

\usepackage{comment}
\usepackage{graphicx}

\usepackage{amsmath,amssymb}
\usepackage{bm}
\usepackage{enumitem}

\usepackage{natbib}
\usepackage{wasysym}

\newcommand{\M}{\mathcal{M}}

\begin{document}

\title{Nuclear equation of state from observations of short gamma-ray burst remnants}

\author{Paul D. Lasky}
	\email{paul.lasky@unimelb.edu.au}
		\affiliation{School of Physics, University of Melbourne, Parkville, VIC 3010, Australia}
\author{Brynmor Haskell}
%	\email{brynmor.haskell@unimelb.edu.au}
	\affiliation{School of Physics, University of Melbourne, Parkville, VIC 3010, Australia}	
\author{Vikram Ravi}
    \altaffiliation{CSIRO Astronomy and Space Science, Australia Telescope National Facility, P.O. Box 76, Epping, NSW 1710, Australia}
	\affiliation{School of Physics, University of Melbourne, Parkville, VIC 3010, Australia}
%	\email{v.vikram.ravi@gmail.com}
%\email{ejhowell@physics.uwa.edu.au}
 \author{Eric J. Howell}
\affiliation{School of Physics, University of Western Australia, Crawley WA 6009, Australia}
\author{David M. Coward}
\affiliation{School of Physics, University of Western Australia, Crawley WA 6009, Australia}

\pacs{
26.60.-c, % Nuclear matter aspects of neutron stars
%26.60.Kp, % Equations of state of neutron-star matter
97.60.Jd, % Neutron stars (see also 26.60.-c Nuclear matter aspects of neutron stars in—Nuclear physics)
04.30.Tv % Gravitational-wave astrophysics (see also 95.85.Sz Gravitational radiation, magnetic fields, and other observations in astronomy)
}

\begin{abstract}
	%Two neutron stars merge, then they wait, then they collapse.  This is definitely interesting.
The favored progenitor model for short $\gamma$-ray bursts (SGRBs) is the merger of two neutron stars that triggers an explosion with a burst of collimated $\gamma$-rays. Following the initial prompt emission, some SGRBs exhibit a plateau phase in their X-ray light curves that indicates additional energy injection from a central engine, believed to be a rapidly rotating, highly magnetized neutron star.  The collapse of this ``protomagnetar" to a black hole is likely to be responsible for a steep decay in X-ray flux observed at the end of the plateau.  In this paper, we show that these observations can be used to effectively constrain the equation of state of dense matter.  In particular, we show that the known distribution of masses in binary neutron star systems, together with fits to the X-ray light curves, provides constraints that exclude the softest and stiffest plausible equations of state.  We further illustrate how a future gravitational wave observation with Advanced LIGO/Virgo can place tight constraints on the equation of state, by adding into the picture a measurement of the chirp mass of the SGRB progenitor.
\end{abstract}

\maketitle

Recent observations of long and short $\gamma$-ray bursts (SGRBs) show plateau phases in the X-ray light curves that last hundreds of seconds \cite{obrien06,nousek06,evans09,rowlinson10a,rowlinson13,gompertz13} and provide evidence for ongoing energy injection through a central engine \cite{nousek06,zhang06,metzger08a}.  The main candidate for the central engine in SGRBs is a rapidly rotating, highly magnetized neutron star (NS) \cite{dai98a,dai98b,zhang01,gao06} that forms following the coalescence of two NSs \cite{eichler89,barthelmy05,berger05,bloom06,duez10a,rezzolla11,Tanvir2013}.  Recent analytic fits to X-ray light curves support this ``protomagnetar'' interpretation of a central engine for both long \cite{troja07,lyons10,dallosso11,bernardini12} and short GRBs \cite{rowlinson10a,rowlinson13,gompertz13}.  Excitingly, some objects exhibit an abrupt cutoff in the X-ray flux $\sim100\,{\rm s}$ after the initial trigger \cite{troja07,lyons10,rowlinson13}.  This has been interpreted as the metastable protomagnetar collapsing to form a black hole.

From a theoretical perspective, the coalescence of binary NSs can follow a number of evolutionary paths.  If the merger remnant is sufficiently massive, it immediately collapses to a black hole, or forms a dynamically unstable hypermassive NS that is supported by strong differential rotation and thermal pressure \cite{rezzolla11,hotokezaka13}.  Magnetic braking terminates differential rotation on the Alfv\'en timescale \cite{baumgarte00,shapiro00}
%which, given the internal magnetic field is wound up to $\sim10^{16}$--$10^{17}\,{\rm G}$ \cite{thompson93,price06}, 
implying that the object collapses in $\sim10$--$100\,{\rm ms}$.  If the merger remnant is less massive it forms a supramassive, metastable protomagnetar \cite{rosswog02,giacomazzo13} in which centrifugal forces from uniform rotation support a higher mass than the nonrotating Tolman-Oppenheimer-Volkoff (TOV) maximum mass \cite{cook94}.  Such a supramassive star spins down until the centrifugal force is insufficient to support the mass, at which point it collapses to a black hole. The recent discovery of $\sim2\,M_\odot$ NSs \cite{demorest10,antoniadis13} demonstrates that the equation of state (EOS) permits massive enough NSs for supramassive stars to be created from the merger of two NSs \cite{hebeler13}. Finally, a merger remnant that is less massive than the TOV maximum mass will survive as a stable NS.

In this paper, we focus on the possibility that protomagnetars drive the plateau phases of SGRB X-ray light curves.  The loss of rotational energy from the NS powers the emission, and a simple spin-down model can be fit to the light curve to obtain the initial spin period, $p_0$, and surface dipolar magnetic field, $B_p$, of the protomagnetar \cite{zhang01,rowlinson10a,rowlinson13}. When an abrupt decay in X-ray luminosity is also observed, this is interpreted as the star having spun down to the point at which centrifugal forces can no longer support its mass against gravity \cite{troja07,lyons10}. The time between the initial prompt emission and the decay, $t_{\rm col}$, is hence interpreted as the collapse time of the protomagnetar.  Given $p_0$ and $B_p$, the time it takes the NS to collapse will depend only on its initial mass and the EOS. We thus have {\it almost} all of the ingredients needed to determine the EOS, with the exception that the initial mass of the NS is not known. In the following, we show how one can constrain the EOS using these observations and the observed distribution of NS masses in binary NS systems \cite{kiziltan10,valentim11,kiziltan13}.  We also show how the EOS constraints will improve given a gravitational wave (GW) measurement of the binary inspiral (i.e., prior to coalescence) with Advanced LIGO and Virgo.  

We focus on the observations presented in \citet{rowlinson13}, in which X-ray plateaus were observed following initial SGRB triggers using {\it Swift}.  The light curves fit the prediction of a protomagnetar that is being spun down through dipole electromagnetic radiation \cite{zhang01} (as noted in \citet{rowlinson13}, this is consistent with the late-time residual spin-down phase being driven by a relativistic magnetar wind \cite{metzger11}), allowing the authors to obtain $p_0$ and $B_{p}$ from the model.

%Of the sample of 28 SGRBs in \cite{rowlinson13}, eleven also exhibit an abrupt decay in the $X$-ray luminosity a time $t_{\rm col}$ after the initial trigger.  This is interpreted as the protomagnetar collapsing to a black hole.  Those that do not show a cut in $X$-ray flux are considered to be stable. 

%The rate at which a star spins down due to electromagnetic dipole radiation is governed by $B_p$, $p_0$, 
%the magnetic field inclination angle\footnote{{\bf we need to do something about this. It's easy enough to add as a parameter in the spin down, claim we don't know it, and work out how big the effect is.}} 
%and structural parameters of the star, i.e., radius, $R$, and moment of inertia, $I$.  These parameters depend on the mass, $M$, and the EOS.  Moreover, the point in the $M$--$p$ plane at which a supramassive star has lost sufficient angular momentum such that centrifugal forces can no longer support its own mass, is also a function of the EOS.  The observations of \citet{rowlinson13} provide {\it almost} the requisite information to determine the EOS, with the exception that the mass of the post-merger remnant is not known.  

\citet{rowlinson13} present data for a number of objects with accurate redshift measurements.  As this is required to determine the rest-frame light curve, and hence $p_0$ and $B_{p}$, we omit any SGRBs for which the redshift is not known.  We are left with four SGRBs that collapse and four that are long-term stable\footnote{There were six SGRBs that are long-term stable and satisfy our criteria; however, two of these (GRBs 050509B and 061201) did not show conclusive fits to the magnetar model and were therefore labelled by Ref. \cite{rowlinson13} as ``possible candidates''.  We omit these in the present analysis, although note that they are consistent with our general conclusions.}, which are presented in Table \ref{tabl:GRB}.  %This gives the GRB identifier; its redshift, $z$; initial spin period, $p_0$; surface magnetic field, $B_p$; and collapse time, $t_{\rm col}$.  These are the only observational ingredients we require in the present work.  

\begin{table}
	\centering
	\caption{\label{tabl:GRB}  The SGRB sample containing central engines used in this article, with all data and fits from Ref. \cite{rowlinson13}. $z$, $p_0$, $B_p$ and $t_{\rm col}$ are, respectively, the redshift, initial spin period, surface dipolar magnetic field, and collapse time. The bottom four SGRBs do not collapse within $10^4$ -- $10^5\,{\rm s}$.}
    \begin{tabular}{|ccccc|}
		\hline\hline
		GRB & $z$ & $p_0$ & $B_p$ & $t_{\rm col}$\\
		& & $\,[\rm ms]$ & $\,[10^{15}\,{\rm G}]$ & $[{\rm s}]$\\
		\hline
		060801 & $1.13$ & $1.95^{+0.15}_{-0.13}$ & $11.24^{+1.93}_{-1.78}$ & $326$ \\
		070724A & $0.46$ & $1.80^{+1.04}_{-0.38}$ & $28.72^{+1.42}_{-1.29}$ & $90$\\
		080905A & $0.122$ & $9.80^{+0.78}_{-0.77}$ & $39.26^{+10.24}_{-12.16}$ & $274$\\
		101219A & $0.718$ & $0.95^{+0.05}_{-0.05}$ & $2.81^{+0.47}_{-0.39}$ & $138$\\
		\hline
		%050509B & $0.23$ & $80.32^{+24.98}_{-17.91}$ & $21.85^{+16.44}_{-11.98}$ & -- \\
		051221A & $0.55$ & $7.79^{+0.31}_{-0.28}$ & $1.80^{+0.14}_{-0.13}$ & --\\
		%061201 & $0.111$ & $14.52^{+0.59}_{-0.52}$ & $19.00^{+1.75}_{-1.44}$ & --\\
		070809 & $0.219$ & $5.54^{+0.48}_{-0.43}$ & $2.06^{+0.48}_{-0.42}$ & --\\
		090426\footnote{Duration ($T90$=1.2s) suggests GRB090426  is a SGRB, however its host and prompt characteristics remain ambiguous \citep[e.g.][]{Lu2010ApJ,Guelbenzu2011AA}.} & $2.6$ & $1.89^{+0.08}_{-0.07}$ & $4.88^{+0.88}_{-0.90}$ & --\\
		090510 & $0.9$ & $1.86^{+0.04}_{-0.03}$ & $5.06^{+0.27}_{-0.23}$ & --\\
		\hline\hline
	\end{tabular}
\end{table}

%\begin{tabular}{|cccc|ccc|}
%    	\hline\hline
%		GRB & $p_0$ & $B_p$ & $t_{\rm col}$ & GRB & $p_0$ & $B_p$ \\
%		& $\,[\rm ms]$ & $\,[10^{16}\,{\rm G}]$ & $[{\rm s}]$ & & $\,[\rm ms]$ & $\,[10^{16}\,{\rm G}]$ \\
%		\hline
%		060801 & $1.95^{+0.15}_{-0.13}$ & $1.12^{+0.19}_{-0.18}$ & $326$ 
%        & 051221A & $7.79^{+0.31}_{-0.28}$ & $0.18^{+0.01}_{-0.01}$\\
%		070724A & $1.80^{+1.04}_{-0.38}$ & $2.87^{+0.14}_{-0.13}$ & $90$
%        & 070809 & $5.54^{+0.48}_{-0.43}$ & $0.21^{+0.05}_{-0.04}$\\
%		080905A & $9.80^{+0.78}_{-0.77}$ & $3.93^{+1.02}_{-1.22}$ & $274$ 
%        & 090426 & $1.89^{+0.08}_{-0.07}$ & $0.49^{+0.09}_{-0.09}$\\
%		101219A & $0.95^{+0.05}_{-0.05}$ & $0.28^{+0.05}_{-0.04}$ & $138$
%        & 090510 & $1.86^{+0.04}_{-0.03}$ & $0.51^{+0.03}_{-0.02}$\\
%		\hline\hline
%	\end{tabular}
%\end{table}

The values of $B_p$ and $p_0$ are derived assuming electromagnetic dipolar spin-down, with perfect efficiency in the conversion between rotational energy and electromagnetic radiation.  We discuss the possibility of a lower efficiency below. Note that a mass of $1.4 M_\odot$ and radius of $10$ km were also assumed, although the dependence on these parameters is weak \cite{rowlinson10a,rowlinson13}. The standard spin-down formula is \cite{shapiro83}
\begin{equation}
	p(t)=p_0\left(1+\frac{4\pi^2}{3c^2}\frac{B_pR^6}{Ip_0^2}t\right)^{1/2},\label{eq:Pt}
\end{equation}
where $R$ and $I$ are the radius and moment of interia, respectively, of the NS.  This spin-down law is implicitly used in the fits to the X-ray light curves \cite{rowlinson13}; a deviation from dipole spin-down would result in a different power-law exponent (see also \cite{zhang01}).  Moreover, it has recently been shown that randomly distributed magnetic fields lead to similar spin-down luminosities than ordered magnetic fields \cite{rezzollaprivate13}.  

For a given EOS, one can write the maximum gravitational mass, $M_{\rm max}$, as a function of the star's rotational kinetic energy \cite{shapiro83,lyford03}, and hence $p$. For slow rotation
%\begin{equation}
%M_{\rm max}=M_{\rm TOV}\left(1-2\frac{T}{\left|W\right|}\right)^{-3/2},\label{eq:Mmax1}
%\end{equation}
%where $T$ and $W$ are the rotational kinetic and potential energies respectively.  Equation (\ref{eq:Mmax1}) is appropriate for rotating stars in Newtonian gravity.  Expanding the right side of (\ref{eq:Mmax1}) for small $T/|W|$, we write
\begin{equation}
	M_{\rm max}=M_{\rm TOV}\left(1+\alpha p^\beta\right),\label{eq:Mmax}
\end{equation}
where in Newtonian gravity $\beta=-2$ and $\alpha$ is a function of the star's mass, radius, and moment of inertia.  We evaluate equation (\ref{eq:Mmax}) in relativistic gravity by creating equilibrium sequences of $M_{\rm max}(p)$ using the general relativistic hydrostatic equilibrium code {\tt RNS} \cite{stergioulas95}.  That is, for various values of the spin period we calculate equilibrium sequences and find the local maximum in the $M$--$\rho_c$ curve (where $\rho_c$ is the central energy density) that indicates the maximum mass.  We then calculate a functional fit to these equilibrium sequences to get $\alpha$ and $\beta$ for each EOS.
%\footnote{\bf should we provide these numbers, including an estimate of the residual error, for each EOS?}.

A supramassive protomagnetar collapses when the star's period becomes large enough that $M_{\rm p}=M_{\rm max}(p)$, where $M_{\rm p}$ is the mass of the protomagnetar.  The collapse time, $t_{\rm col}$, is found by substituting (\ref{eq:Pt}) into (\ref{eq:Mmax}) with $t=t_{\rm col}$ and $M_{\rm max}=M_{\rm p}$.  Solving for $t_{\rm col}$ gives
\begin{equation}
	t_{\rm col}=\frac{3c^3I}{4\pi^2 B_p^2R^6}\left[\left(\frac{M_{\rm p}-M_{\rm TOV}}{\alpha\,M_{\rm TOV}}\right)^{2/\beta}-p_0^2\right].\label{eq:tcol}
\end{equation}
Equation (\ref{eq:tcol}) gives the time for a supramassive protomagnetar to collapse to a black hole given observed parameters ($p_0$, $B_p$, $M_{\rm p}$) and parameters related to the EOS ($M_{\rm TOV}$, $R$, and $I$). Note that Eq. (\ref{eq:tcol}) does not account for several effects, such as how $I$ and $B_p$ change with time as the star spins down or how the presence of matter outside the star affects the spin-down torque, a point we discuss below.
%Nevertheless it allows us a reasonably simple estimate of the parameters to prove the potential of our method.

The observations in Ref. \cite{rowlinson13} give $B_p$, $p_0$ and $t_{\rm col}$, implying that we require $M_{\rm p}$ in Eq. (\ref{eq:tcol}) to constrain the EOS.  We obtain $M_{\rm p}$ statistically from the observed masses of NSs in binary NS systems \cite{kiziltan10,valentim11,kiziltan13}, where the most up-to-date measurements give $M=1.32^{+0.11}_{-0.11}\,M_\odot$, with the errors being the $68\%$ posterior predictive intervals \cite{kiziltan13}.  Numerical simulations of binary NS mergers and observations of SGRBs indicate that $\lesssim0.01\,M_\odot$ of material is ejected during the merger \cite[e.g.,][and references therein]{hotokezaka13,giacomazzo13b}.  Modulo this lost mass, which we ignore in the following, it is the rest mass of a system that is conserved through the merger.  An approximate conversion between gravitational and rest masses is $M_{\rm rest}=M+0.075M^2$ \cite{timmes96}, which leads to a gravitational mass for the protomagnetar following an SGRB merger of $M_{\rm p}=2.46^{+0.13}_{-0.15}\,M_{\odot}$.

\begin{figure*}
	\begin{center}
		\includegraphics[angle=0,width=1.0\textwidth]{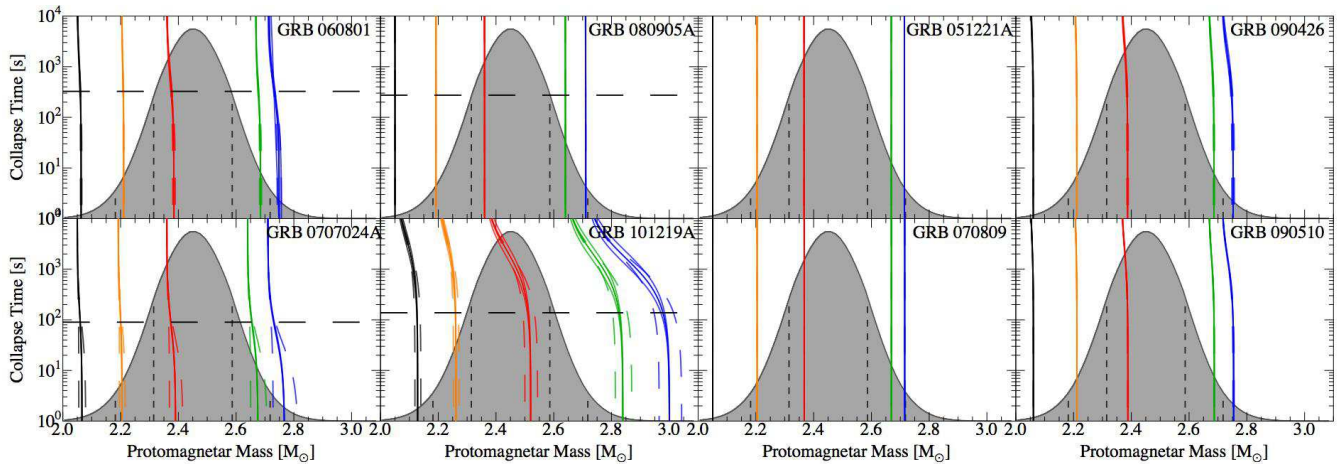}
	\end{center}
	\caption{\label{fig:tcolM}Collapse time as a function of the protomagnetar mass for each of the SGRBs in Table \ref{tabl:GRB}.  The two left-hand columns are those in which the protomagnetar collapses to form a black hole, where the collapse time is given by the horizontal dashed black line.  The two right-hand columns are those SGRBs that form stable protomagnetars.  The theoretical collapse time for each EOS is calculated from Eq. (\ref{eq:tcol}), where the initial spin period and magnetic field distributions are given in Table \ref{tabl:GRB} for each GRB.  Five EOSs are shown in each panel:  SLy (black), APR (orange), GM1 (red), AB-N (green) and AB-L (blue).  The dark solid curve for each EOS assumes the values of $p_0$ and $B_p$ given in Table \ref{tabl:GRB} with the 68\% confidence intervals in $p_0$ and $B_p$ included in the faded dashed and faded solid curves respectively.  The shaded region is the protomagnetar mass distribution that results from merging two NSs whose masses are independently drawn from the binary NS mass distribution of Ref. \cite{kiziltan13}, with the $68\% $ and $95\%$ mass intervals represented with the vertical dashed lines.
	}
\end{figure*}

In Fig. \ref{fig:tcolM}, we plot the collapse time, $t_{\rm col}$, as a function of the  protomagnetar mass, $M_{\rm p}$, for each of the SGRBs listed in Table \ref{tabl:GRB}.  We utilize five EOSs that are consistent with current observations and have a range of maximum masses: SLy \cite{SLy} ($M_{\rm TOV}=2.05\,M_\odot$, $R=9.97\,{\rm km}$; black curve), APR \cite{APR} ($2.20\,M_\odot$, $10.00\,{\rm km}$; orange), GM1 \cite{GM1} ($2.37\,M_\odot$, $12.05\,{\rm km}$; red), AB-N \cite{arnett77} ($2.67\,M_\odot$, $12.90\,{\rm km}$; green) and AB-L \cite{arnett77} ($2.71\,M_\odot$, $13.70\,{\rm km}$; blue).  
%For each EOS, the dark solid curve assumes the values of $p_0$ and $B_p$ in Table \ref{tabl:GRB}, with the errors in $p_0$ and $B_p$ included in the faded dashed and faded solid curves respectively.  
%The two left-hand columns of Figure \ref{fig:tcolM} show the four SGRBs that collapse to form a black hole in time $t_{\rm col}$, which is indicated by the horizontal dashed line in these figures.  The two right-hand columns are those that do not show evidence of collapse.  Finally, the shaded region is the protomagnetar mass distribution derived above -- i.e., with $M_{\rm p}=2.46^{+0.13}_{-0.15}\,M_\odot$, with the vertical dashed lines showing the $68\%$ and $95\%$ posterior predictive intervals.

Consider GRB 060801 in Fig. \ref{fig:tcolM}, with $B_p$ and $p_0$ given in Table \ref{tabl:GRB}.  EOS GM1 (red curves) requires $M_{\rm p}\approx2.38\,M_{\odot}$ for it to collapse $326\,{\rm s}$ following the initial burst.  On the other hand, EOS SLy (black curves) requires $M_{\rm p}\approx2.06\,M_{\odot}$, which falls well outside the $2\sigma$ posterior mass distribution.  The quoted errors for $B_p$ and $p_0$ have little effect on this result.  Similarly, GRB 101219A requires $M_{\rm p}\approx\,3.00M_\odot$ for AB-L and $M_{\rm p}\approx2.82\,M_\odot$ for AB-N, which both lie at the extreme high-mass end of the distribution.  In this sense, all of the GRBs plotted in the two left-hand columns of Fig. \ref{fig:tcolM} favor the intermediate EOSs.  It is worth noting that the EOSs we plot are a representative sample that covers a wide range of maximum masses; many more EOSs fit into the intermediate regime that would be satisfied by the constraints we are placing herein.  For an up-to-date review of plausible EOSs see Ref. \cite{hebeler13}. 

It is worth paying special attention to GRB 080905A.  \citet{rowlinson13} found relatively large $p_0$, implying slow spin-down from electromagnetic torques.  In the $274\,{\rm s}$ before GRB 080905A collapses, the protomagnetar has spun down from $p_0=9.8\,{\rm ms}$ to between $p=10.2\,{\rm ms}$ and $p=10.9\,{\rm ms}$ depending on the EOS.  For any EOS, this requires a fine tuning in the protomagnetar mass. %on the order of one part in $10^{5}$, as calculated by looking at the change in the maximum mass between the initial and final spin periods\footnote{\bf Need to think about the right way to quantify this.}.  %In terms of figure \ref{fig:tcolM}, this appears as each curve for $t_{\rm col}(M_{\rm p})$ appearing as an almost vertical line.
There are many interpretations for this fine tuning.  \citet{fan13} proposed that this is evidence that the protomagnetar was predominantly spun down through GW losses as opposed to electromagnetic torques.  This is possible, although we note that the ellipticity of the star needs to be $\sim 10^{-2}$--$10^{-3}$, which requires an average internal toroidal field of almost $10^{\rm 17}\,{\rm G}$ for a star with $M\gtrsim2.5\,M_\odot$ \cite{haskell08}. On the other hand, the isotropic efficiency of turning rotational energy into electromagnetic energy is assumed to be $100\%$.  Reducing the assumed efficiency or beam opening angle also leads to a reduction of the initial spin period.  Other possibilities include a chance	alignment that led to a false host-galaxy identification, or	ongoing accretion or propellering that is affecting the pulsar spin-down \cite{rowlinson13b}.  It is clear that these are crucial issues that have to be dealt with in a more systematic study if our method is to be used to obtain a strong, quantitative constraint on the EOS.

The two right-hand columns of Fig. \ref{fig:tcolM} are those SGRBs that are not observed to collapse.  Their relatively high initial spin periods and low surface magnetic field strengths imply that they do not spin down significantly in $\sim100$--$1000\,{\rm s}$.  Therefore, as with GRB 080905A, each EOS curve is almost a vertical line.  If $M_{\rm p}\le M_{\rm TOV}$, these objects are stable magnetars, and will never collapse from loss of centrifugal support.  On the other hand, they may still have $M_{\rm p}\gtrsim M_{\rm TOV}$, in which case they are unstable with $t_{\rm col}\gg10^{5}\,\rm{s}$.  If the latter is true, these GRBs could be candidate ``blitzars'' \cite{falcke13} that are a proposed physical mechanism behind fast radio bursts (FRBs) \cite{lorimer07,thornton13}.  In principal, if the blitzar model is correct one could utilize the method described herein to also constrain the EOS using FRBs, although a 
%GRB remnant\footnote{\bf what's the right wording here} would be required to determine $t_{\rm col}$.
method for determining $t_{\rm col}$ would be required.
A method for testing the blitzar model, in particular the connection between FRBs and GRBs, has recently been described in Ref. \cite{zhang13}.

\begin{figure}
	\begin{center}
		\includegraphics[angle=0,width=0.92\columnwidth]{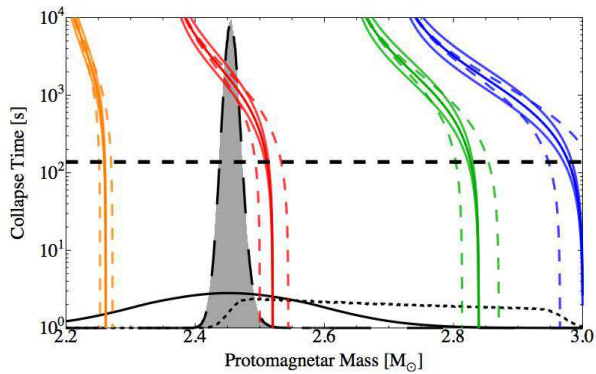}
	\end{center}
	\caption{\label{fig:tcolMAllDists}Collapse time as a function of the protomagnetar mass for GRB 101219A.  The five equations of state are described in the caption of Fig. \ref{fig:tcolM}.  The three protomagnetar mass distributions represent the binary NS mass distribution only (solid black; same as the distribution shown in Fig. \ref{fig:tcolM}), the posterior mass distribution from a conservative Advanced LIGO/Virgo measurement of the progenitor chirp mass and symmetric mass ratio (dotted black) and the posterior mass distribution from the binary NS distribution and the conservative Advanced LIGO/Virgo progenitor measurement (dashed black with shading).
	}
\end{figure}

Figure \ref{fig:tcolM} shows what currently can be achieved given that the mass of the GRB remnant can only be statistically inferred from binary NS observations.  In the near future, Advanced LIGO and Virgo will begin measuring GWs from binary NS inspirals at a rate of $0.4$ to $400$ per year \cite{abadie10}.  Importantly for our purposes, these instruments will measure the chirp mass, $\M=(m_1m_2)^{3/5}/\left(m_1+m_2\right)^{-1/5}$, and the symmetric mass ratio $\eta=(m_1m_2)/(m_1+m_2)^2$, where $m_{1,2}$ are the masses of the original progenitor NSs. The fractional 95\% confidence intervals for these quantities will, at worst, be $\sim2\%$ for $\M$ and $\sim20\%$ for $\eta$ (see Ref. \cite{aasi13} for details, which includes an exhaustive discussion of data-analysis algorithms for parameter estimation from GW measurements). These measurements will allow $m_{1}$ and $m_{2}$ to be estimated.  In Fig. \ref{fig:tcolMAllDists} we again plot the collapse time as a function of protomagnetar mass for GRB 101219A, but assume a hypothetical GW measurement of the merger of two $1.32\,M_\odot$ NSs with the aforementioned confidence intervals for $\M$ and $\eta$. 

Figure \ref{fig:tcolMAllDists} clearly shows that a combined measurement of the NS progenitor masses using GWs and knowledge of the prior NS mass distribution significantly tightens the constraints on the EOS. For example, assuming the GM1 EOS (red curve in Figure \ref{fig:tcolMAllDists}) implies a 
protomagnetar mass for the remnant of GRB 101219A of $M_{\rm p}=2.51^{+0.02}_{-0.02}\,M_{\odot}$ (68\% confidence interval). While this is broadly consistent with the expectation of $M_{\rm p}=2.45^{+0.14}_{-0.14}\,M_{\odot}$ from current constraints on the binary NS mass distribution (see Fig. \ref{fig:tcolM}), this would be inconsistent with a putative Advanced LIGO measurement of $\M$ for the merger of two $1.32\,M_\odot$ NSs combined with binary NS mass distribution constraints. The latter would imply $M_{\rm p}=2.46^{+0.02}_{-0.02}\,M_{\odot}$. 
%$M_{\rm p}=2.456^{+0.016}_{-0.016}\,M_{\odot}$. 

The apparent bias in the protomagnetar mass posterior distribution from Advanced LIGO-only measurements (dotted black curve of Fig. \ref{fig:tcolMAllDists}) is caused by the bias in the estimation of $\eta$ for some GW waveform templates \cite{aasi13}. Individual templates can, however, be used to estimate $\eta$ with percent-level precision, which, when combined with measurements of $\M$, would render the binary NS mass distributions irrelevant in constraining protomagnetar masses.

%For the GM1 EOS used in the red curve of Figure \ref{fig:tcolMAllDists}, we find the combined mass is $M_{\rm p}=2.52^{+0.44}_{-0.44}\,M_{\odot}$, implying the $68\%$ confidence level has an error in $M_{\rm p}$ on the order of $1.8\%$ compared to $5.3\%$ for the neutron star mass distribution only.  
%Interestingly, the gravitational wave plus NS distribution in figure \ref{fig:tcolMAllDists} does not change when only the chirp mass is measured.  {\bf we should quantify and explain this with one more sentence.}  {\bf Vikram: We also need a sentence or two here (or slightly above) on the distribution for a LIGO only measurement.}

How often does one expect a coincident GW and electromagnetic detection of an SGRB with an X-ray plateau?  Using a conservative beaming angle of $8^\circ$ \cite[][ and references therein]{Tanvir2013,bartos13} and the \emph{Swift} sample of SGRBs corrected for dominant selection biases \cite{coward12}, we obtain an intrinsic rate of $820\,{\rm Gpc}^{-3}\,{\rm s}^{-1}$.  With a binary NS horizon distance for coincident Advanced LIGO and Virgo detections \cite{abadie10,lsc_2013} and assuming $50\%$ of all SGRBs have X-ray plateaus \cite{rowlinson13}, we get a rate of 0.2 coincident electromagnetic and GW detections per year.  The Space-based multi-band astronomical Variable Object Monitor (SVOM) has a decrease in sensitivity of a factor $\sim2$ compared to \emph{Swift}, but the higher triggering energy band may be more optimal for the detection of spectrally harder SGRBs.  Assuming that these two effects cancel, the increased sky coverage of SVOM over \emph{Swift} implies $\sim0.4$ coincident events per year.  Finally, ISS-Lobster, a proposed all-sky X-ray imaging telescope, has been estimated to see about two coincident SGRBs per year \cite{Camp2013}, corresponding to about one per year with X-ray plateaus.

In this paper we have shown how one can constrain the nuclear EOS from observations of SGRBs that exhibit X-ray plateaus. We have outlined how current understanding of the mass distribution in NS binaries can already be used to place constraints on the EOS of dense matter and how a future coincident detection of a GW and X-ray signal from a binary NS merger could place significantly stronger constraints. This is an exciting prospect, and the rates we have estimated for coincident detection suggest that it is a very real possibility.

Given this encouraging starting point, it is crucial for future work to build on the method presented here and address in a more systematic way the caveats we have mentioned above (e.g., more detailed torque modeling, more accurate fits to light curves).  This has the potential to allow for strong and truly quantitative constraints to be placed on the EOS of dense matter.

%There are many caveats to our model that consider further investigation before a rigorous constraint can be claimed.  For example, a wrong assumption about the efficiency of conversion from rotational energy into electromagnetic emission changes the inferred initial spin period, and hence the shape of each of the curves in figures \ref{fig:tcolM} and \ref{fig:tcolMAllDists}.  We have further assumed that the spin-down of the protomagnetar is dominated by electromagnetic dipole torques, stating that these dominate over the spin-down due to gravitational wave emission (in contrast to Ref. \cite{fan13}).  This is based on the assumption that the internal magnetic field does not induce a stellar ellipticity on the order of $10^{-2}$--$10^{-3} $ ({\bf check numbers}), which requires an average internal toroidal field of almost $10^{\rm 17}\,{\rm G}$ for a star with $M\gtrsim2.5\,M_\odot$ \cite{cutler02,stella05,dallosso09}.  

\vspace{-0.35cm}

%We should comment on the GW signature of supramassive star collapse.

%\acknowledgements
%We are indebted to David Coward, Andrew Melatos, Mark Bennett and Linqing Wen for valuable discussions.  
\section{Acknowledgments}
We are extremely grateful to Antonia Rowlinson, Luciano Rezzolla and Wen-fai Fong for valuable comments.  We also gratefully acknowledge Jordan Camp who carefully read the manuscript as part of a LIGO Scientific Collaboration review (LIGO document number P1300195).  P.D.L. is supported by the Australian Research Council (ARC) Discovery Project (DP110103347) and an internal University of Melbourne Early Career Researcher grant.  E.J.H. acknowledges support from a UWA Research Fellowship. D.M.C. is supported by an ARC Future Fellowship and B.H. by an ARC DECRA Fellowship.  V.R. is a recipient of a John Stocker Postgraduate Scholarship from the Science and Industry Endowment Fund. This work was made possible through a UWA Research Collaboration Award.

\bibliography{Eos_from_SGRBs}

\end{document}